\documentstyle[11pt]{article}
\newcounter{subequation}[equation]

\makeatletter

\def\thesubequation{\theequation\@alph\c@subequation}
\def\@subeqnnum{{\rm (\thesubequation)}}
\def\slabel#1{\@bsphack\if@filesw {\let\thepage\relax
   \xdef\@gtempa{\write\@auxout{\string
      \newlabel{#1}{{\thesubequation}{\thepage}}}}}\@gtempa
   \if@nobreak \ifvmode\nobreak\fi\fi\fi\@esphack}
\def\subeqnarray{\stepcounter{equation}
\let\@currentlabel=\theequation\global\c@subequation\@ne
\global\@eqnswtrue
\global\@eqcnt\z@\tabskip\@centering\let\\=\@subeqncr
$$\halign to \displaywidth\bgroup\@eqnsel\hskip\@centering
  $\displaystyle\tabskip\z@{##}$&\global\@eqcnt\@ne
  \hskip 2\arraycolsep \hfil${##}$\hfil
  &\global\@eqcnt\tw@ \hskip 2\arraycolsep
  $\displaystyle\tabskip\z@{##}$\hfil
   \tabskip\@centering&\llap{##}\tabskip\z@\cr}
\def\endsubeqnarray{\@@subeqncr\egroup
                     $$\global\@ignoretrue}
\def\@subeqncr{{\ifnum0=`}\fi\@ifstar{\global\@eqpen\@M
    \@ysubeqncr}{\global\@eqpen\interdisplaylinepenalty \@ysubeqncr}}
\def\@ysubeqncr{\@ifnextchar [{\@xsubeqncr}{\@xsubeqncr[\z@]}}
\def\@xsubeqncr[#1]{\ifnum0=`{\fi}\@@subeqncr
   \noalign{\penalty\@eqpen\vskip\jot\vskip #1\relax}}
\def\@@subeqncr{\let\@tempa\relax
    \ifcase\@eqcnt \def\@tempa{& & &}\or \def\@tempa{& &}
      \else \def\@tempa{&}\fi
     \@tempa \if@eqnsw\@subeqnnum\refstepcounter{subequation}\fi
     \global\@eqnswtrue\global\@eqcnt\z@\cr}
\let\@ssubeqncr=\@subeqncr
\@namedef{subeqnarray*}{\def\@subeqncr{\nonumber\@ssubeqncr}\subeqnarray}
\@namedef{endsubeqnarray*}{\global\advance\c@equation\m@ne%
                           \nonumber\endsubeqnarray}

\makeatletter
\@addtoreset{equation}{section}
\makeatother
\renewcommand{\theequation}{\thesection.\arabic{equation}}

\def\dalemb#1#2{{\vbox{\hrule height .#2pt
        \hbox{\vrule width.#2pt height#1pt \kern#1pt
                \vrule width.#2pt}
        \hrule height.#2pt}}}

\let\a=\alpha    \let\e=\epsilon
  \let\q=\theta  
  \let\n=\nu

\def\nn{\nonumber} \def\bd{\begin{document}} \def\ed{\end{document}}
\def\ds{\documentstyle} \let\fr=\frac \let\bl=\bigl \let\br=\bigr
\let\Br=\Bigr \let\Bl=\Bigl
\let\bm=\bibitem
\let\na=\nabla
\let\pa=\partial \let\ov=\overline
\def\ie{{\it i.e.\ }}
\newcommand{\be}{\begin{equation}}
\newcommand{\ee}{\end{equation}}
\def\ba{\begin{array}}
\def\ea{\end{array}}
\def\ft#1#2{{\textstyle{{\scriptstyle #1}\over {\scriptstyle #2}}}}
\def\fft#1#2{{#1 \over #2}}
\def\del{\partial}
\def\sst#1{{\scriptscriptstyle #1}}
\def\oneone{\rlap 1\mkern4mu{\rm l}}
\def\e7{E_{7(+7)}}
\def\td{\tilde}
\def\wtd{\widetilde}
\def\im{{\rm i}}
\def\bog{Bogomol'nyi\ }
\def\q{{\tilde q}}
\def\hast{{\hat\ast}}
\def\0{{\sst{(0)}}}
\def\1{{\sst{(1)}}}
\def\2{{\sst{(2)}}}
\def\3{{\sst{(3)}}}
\def\4{{\sst{(4)}}}
\def\5{{\sst{(5)}}}
\def\6{{\sst{(6)}}}
\def\7{{\sst{(7)}}}
\def\8{{\sst{(8)}}}
\def\n{{\sst{(n)}}}
\def\oo{{\"o}}
\def\hA{\hat{\cal A}}
\def\ns{{\sst {\rm NS}}}
\def\rr{{\sst {\rm RR}}}
\def\tH{{\widetilde H}}
\def\tB{{\widetilde B}}
\def\cA{{\cal A}}
\def\cF{{\cal F}}
\def\tF{{\wtd F}}
\def\Z{\rlap{\sf Z}\mkern3mu{\sf Z}}
\def\ep{{\epsilon}}
\def\IIA{{\rm IIA}}
\def\IIB{{\rm IIB}}
\def\ads{{\rm AdS}}
\def\R{\rlap{\rm I}\mkern3mu{\rm R}}

\def\Ei{{\hbox{Ei}}}
\def\Ci{{\hbox{Ci}}}
\def\Si{{\hbox{Si}}}

\newcommand{\ho}[1]{$\, ^{#1}$}
\newcommand{\hoch}[1]{$\, ^{#1}$}
\newcommand{\bea}{\begin{eqnarray}}
\newcommand{\eea}{\end{eqnarray}}
\newcommand{\ra}{\rightarrow}
\newcommand{\lra}{\longrightarrow}
\newcommand{\Lra}{\Leftrightarrow}
\newcommand{\ap}{\alpha^\prime}
\newcommand{\bp}{\tilde \beta^\prime}
\newcommand{\tr}{{\rm tr} }
\newcommand{\Tr}{{\rm Tr} }
\newcommand{\NP}{Nucl. Phys. }
\newcommand{\tamphys}{\it Center for Theoretical Physics,
Texas A\&M University, College Station, TX 77843}
\newcommand{\upenn}{\it Dept. of Phys. and Astro.,
University of Pennsylvania,
Philadelphia, PA 19104}

\newcommand{\auth}{M. Cveti\v{c}\hoch{\dagger1},
James T. Liu\hoch{\star2}, H. L\"u\hoch{\dagger1} and
C.N. Pope\hoch{\ddagger3} }

\begin{document}
\begin{flushright}
\hfill{
UPR/0847-T \ \ \ \ 
CTP TAMU-19/99 \ \ \ \ 
RU99-6-B\ \ \ \ \  May 1999}\\
\hfill{\bf hep-th/9905096}\\
\end{flushright}


\begin{center}

{\large {\bf Domain-wall Supergravities from Sphere Reduction}}

\vspace{20pt}

\auth

\vspace{10pt}
{\hoch{\dagger}\upenn}

\vspace{10pt}
{\hoch{\ddagger}\tamphys}

\vspace{10pt}
{\hoch{\star}{\it Department of Physics, The Rockefeller University,
New York, NY 10021} }

\vspace{30pt}

\underline{ABSTRACT}
\end{center}

             Kaluza-Klein sphere reductions of supergravities that
admit AdS$\times$Sphere vacuum solutions are believed to be
consistent.  The examples include the $S^4$ and $S^7$ reductions of
eleven-dimensional supergravity, and the $S^5$ reduction of
ten-dimensional type IIB supergravity.  In this paper we provide
evidence that sphere reductions of supergravities that admit instead
Domain-wall$\times$Sphere vacuum solutions are also consistent, where the
background can be viewed as the near-horizon structure of a dilatonic
$p$-brane of the theory. The resulting lower-dimensional theory is a
gauged supergravity that admits a domain wall, rather than AdS, as a
vacuum solution.  We illustrate this consistency by taking the
singular limits of certain modulus parameters, for which the original
$S^n$ compactifying spheres ($n=4,5$ or 7) become $S^p\times R^q$,
with $p=n-q<n$. The consistency of the $S^4$, $S^7$ and $S^5$ reductions
then implies the consistency of the $S^p$ reductions of the
lower-dimensional supergravities.  In particular, we obtain explicit
non-linear ans\"atze for the $S^3$ reduction of type IIA and
heterotic supergravities, restricting to the $U(1)^2$ subgroup of the
$SO(4)$ gauge group of $S^3$.  We also study the black hole solutions
in the lower-dimensional gauged supergravities with domain-wall
backgrounds.  We find new domain-wall black holes which
are not the singular-modulus limits of the AdS black holes of the
original theories, and we obtain their Killing spinors.

{\vfill\leftline{}\vfill
\vskip 10pt \footnoterule
{\footnotesize \hoch{1}
Research supported in part by DOE grant DOE-FG02-95ER40893
\vskip  -12pt} \vskip   14pt
{\footnotesize \hoch{2}
Research supported in part by DOE grant DOE-91ER40651-TASKB
\vskip -12pt} \vskip 14pt
{\footnotesize \hoch{3}
Research supported in part by DOE grant DOE-FG03-95ER40917
\vskip -12pt}  \vskip  14pt
}

\pagebreak
\setcounter{page}{1}

\tableofcontents
\addtocontents{toc}{\protect\setcounter{tocdepth}{2}}
\newpage

\section{Introduction\label{sec:intro}}

    Kaluza-Klein sphere reductions of supergravities that admit
AdS$\times$Sphere vacuum solutions are believed to be consistent.  The
examples include the $S^7$ \cite{duffpope3,DNP} and $S^4$ \cite{PTV}
reductions of eleven-dimensional supergravity, and the $S^5$
\cite{Schwarz,GM,KRV} reduction of ten-dimensional type IIB
supergravity.  The consistency is important since one can then be
assured that the solutions of the resulting lower dimensional gauged
supergravities are also solutions of the M-theory or type IIB theory.
These solutions of the gauged supergravities then provide one side
of the picture of the duality \cite{Maldacena,Gubserklebanovpolyakov,%
Witten,Witten2} between Anti-de Sitter space and conformal field
theories on its boundary.  The complete proof of consistency is still
lacking, owing to the complexity of the reduction ans\"atze.   The
$S^7$ reduction of eleven-dimensional supergravity is better
understood than the other two cases, as a consequence of its having
received more attention owing to its connection with four dimensions.
The full non-linear reduction ansatz, although highly
implicit, was given in \cite{deWitnicolaiwarner,deWitnicolai}.
Recently, explicit non-linear reduction ans\"atze that focus on the
Cartan subgroups of the full gauge groups were obtained for these
three examples \cite{ten}.  The construction shows that the reduction
must be performed at the level of the equations of motion, rather than
in the Lagrangian (even when it exists).  Furthermore, the consistency
of the reduction depends on a delicate balance between the contributions
of the metric and the relevant antisymmetric tensor field strength in the
higher-dimensional supergravity theory.  This conspiracy between the
higher-dimensional fields is equivalent to the one that governs the
supersymmetry of the theory.  

Note that by far the most involved part
of establishing the consistency of any sphere-reduction ansatz is
concerned with the contributions of scalar fields that parameterise
inhomogeneous deformations of the sphere metric.  Specifically, it is
for these inhomogeneous deformations that the conspiracies between terms
in the higher-dimensional Lagrangian are needed in order to achieve 
consistency of the reduction.  This is in contrast to the situation 
in the reduction on a group manifold $G$, as described in \cite{schsch}, where
the scalars parameterise only homogeneous deformations, and the 
consistency of the reduction is guaranteed for each individual term
in the higher-dimensional Lagrangian, by virtue of the invariance of 
the reduction ansatz under the right action of the group $G$.

The consistency of the above sphere reductions raises the question as to
whether it is also possible to perform
consistent Kaluza-Klein sphere reductions on other supergravity
theories that do not admit AdS$\times$Sphere solutions.
AdS$\times$Sphere spacetimes can arise as the near-horizon structures
of the M2-brane, M5-brane and D3-brane.  In these examples there are no
couplings to scalar fields.
However for a generic $p$-brane that couples to a dilaton, the
near-horizon structure of the metric is instead a product of a domain
wall%
\footnote{We use the term ``domain wall'' to mean a $(D-2)$-brane
in $D$ dimensions.}
and a sphere, with a certain warp factor in the
metric.  (In this paper we shall refer to this spacetime as a
Domain-wall$\times$Sphere, even though the warp-factor implies that the
spacetime metric is not a direct product.)  Note that super Yang-Mills
field theories do not only arise at the boundary of AdS space, but
also emerge as the world-volume theories of D-branes.  A generic
D-brane has a near-horizon structure which is a
Domain-wall$\times$Sphere. This raises the question as to whether
supergravity in the D-brane near-horizon background has anything to do
with the super Yang-Mills theory on the D-brane world-volume.  The
correspondence of supergravity on a domain-wall background with
quantum field theory on the wall was proposed in \cite{BST}.

In this paper, we shall address the following question: Is it consistent
to dimensionally reduce the theory with a domain wall$\times$Sphere
solution on the associated sphere?  If so, what will be the
lower-dimensional theory?
In the AdS$\times$Sphere compactification, the lower-dimensional
theory is a gauged supergravity that allows an AdS spacetime as a
vacuum solution.  In other words, the scalar potential of the
spherically-reduced theory has at least one stationary point.
We refer to these theories as ``AdS supergravities.''
In the new cases that we wish to consider here, we would
expect that the lower-dimensional theory would instead admit a domain wall,
rather than an AdS spacetime, as its vacuum solution.
We refer to such theories as ``Domain-wall supergravities.''
Before we address
the question of the consistency of such a reduction, we
shall first examine whether there exist lower dimensional theories
that admit such domain-wall solutions.

        Let us consider gauged $D=7$, $N=2$ supergravity as an
example.  The bosonic sector of the supergravity multiplet contains
the metric, a dilaton
a 3-form vector potential and $SU(2)$ Yang-Mills gauge fields.  The
Lagrangian has the following form \cite{PPV,TV}
\bea
e^{-1}{\cal L}_7 &=& R - \ft12 (\del\phi)^2 - \ft12 m^2\,
e^{\ft{8}{\sqrt{10}} \phi} + 4 g\, m\,
e^{\ft3{\sqrt{10}} \phi} +
4 g^2\, e^{-\ft{2}{\sqrt{10}}\phi}\ ,\nn\\
&& -\ft1{48}e^{-\ft{4}{\sqrt{10}} \phi}\,  (F_\4)^2 -
\ft1{4} e^{\ft2{\sqrt{10}}\phi}\, (F^a_\2)^2\nn\\
&& + \ft12F_\4\wedge F^a_\2\wedge A^a_\1 + \ft12m\, F_\4\wedge A_\3\ .
\label{d7gauge}
\eea
Note that the 3-form gauge potential has a topological mass term, with
mass parameter $m$. The scalar potential has a supersymmetric maximum,
and a non-supersymmetric minimum. The scalar sector admits a
supersymmetric domain-wall solution, given by \cite{lpss}
\bea
ds_{7}^2 &=& e^{2A}\, dx^\mu dx^\nu\eta_{\mu\nu} +
e^{8A}\, dy^2\ ,\qquad e^{-\ft3{\sqrt{10}}\phi} = H\ ,\nn\\
e^{-4A} &=& \fft{8m}{5 (H^{4/3})'} + \fft{2 g}{5\,
(H^{-1/3})'} \ ,\qquad H = e^{-\ft{3}{\sqrt{10}}\phi_0} + q\, |y|\ ,
\eea
where a prime denotes a derivative with respect to $y$.  For generic
non-vanishing $m$ and $g$, we can obtain an AdS$_7$ spacetime by
sending $q$ to zero.  In either of the special cases where $m$=0 or
$g=0$, the domain wall has no AdS$_7$ limit.

     Let us first examine the case when $g=0$.  It was observed in
\cite{lpdomain} that the corresponding domain wall can be obtained
from the vertical dimensional reduction of an M5-brane, whose charges
are uniformly distributed over four of the transverse dimensions,
which are taken as the compactification coordinates.  The
eleven-dimensional 5-brane has the form
\bea
ds_{11}^2 &=& H^{-1/3} dx^{\mu} dx^\nu\eta_{\mu\nu} +
            H^{2/3} (dy^2 + ds_4^2)\ ,\nn\\
F_4&=& q\, \epsilon_4
\eea
where $ds_4^2$ is the metric of a four-dimensional flat space, with
$\epsilon_4$ being its volume form.  Thus the $g=0$ limit can be
viewed as M-theory compactified on a 4-dimensional flat space.  The
topological mass term of the 3-form gauge potential has its origin
in the $FFA$ term of eleven-dimensional supergravity.

           The $m=0$ limit is quite different.  We find that its
higher-dimensional origin is from the near-horizon limit of the
NS-5brane in $D=10$, which has the metric
\be
ds_{10}^2 = H^{-1/4} dx^\mu dx^\nu\eta_{\mu\nu} +
H^{3/4} (dr^2 + r^2\, d\Omega_3^2) \ ,
\ee
with $H = 1 + Q/r^2$.  In the near-horizon limit, we shall have
$H \rightarrow Q/r^2$.

         The seven-dimensional theory (\ref{d7gauge}) can be obtained
from the $S^4$ reduction of M-theory.  With the parameters $g$ and
$m$ both non-zero, it allows AdS$_7$ as a vacuum solution.  In one
singular limit, where $g=0$, it can be obtained from M-theory on a
4-dimensional flat background, with the 4-form field strength having a
term proportional to the volume form of the internal space.  Such a
reduction is indeed consistent, giving rise to a 3-form gauge
potential with a topological mass term \cite{lpdomain}.  The vacuum
domain-wall solution in $D=7$ has its origin as a 5-brane in $D=11$,
with its charges uniformly distributed over the internal space.  In
the other singular limit, with $m=0$, the vacuum domain-wall solution
has its origin instead as the $S^3$ reduction of the NS-5brane in
$D=10$.

       The above example illustrates that there should exist two
different singular limits of the $S^4$ reduction of M-theory, in one
of which the internal 4-space becomes flat, while in the other the
internal space limits to 4-space that contains a 3-sphere factor.  The
consistency of the $S^4$ reduction of M-theory thus implies the
consistency of the $S^3$ reduction of the ten-dimensional theory.

        The main purpose of this paper is to examine the above
conjecture in detail.  In section 2, we show that there exist singular
limits in certain modulus parameters such that $S^4$ becomes either a
flat four-space $\R^4$, or else $S^3\times \R$.  In the flat-space limit, the
M-theory becomes a massive supergravity in $D=7$ with a topologically
massive 3-form potential.  In the $S^3\times \R$ limit, the M-theory
becomes a gauged supergravity, but with no massive 3-form potential.
We also obtain new domain-wall black holes in these
limiting cases.   These are analogous to the AdS black holes of
gauged anti-de Sitter supergravities, but with asymptotic structures that
approach domain walls rather than AdS spacetimes.

         In section 3, we study the $S^7$ reduction of M-theory, and
show that there exist singular limits in certain modulus parameters
such that $S^7$ becomes $S^3\times \R^4$ or $S^5\times \R^2$.  We also
obtain supersymmetric domain-wall black holes in the original theory
before taking the limit.  We study how the solutions behave under
these limiting procedures.  In section 4, we turn out attention to the
$S^5$ reduction of the type IIB theory, and we find a singular limit
where $S^5$ becomes $S^3\times \R^2$.  We also obtain new
supersymmetric domain-wall solutions in the limiting theory.  In the
appendix, we obtain a general class of domain-wall black holes.

\section{$S^4$ reduction of M-theory, and its limits}

       In this section, we show that there exist singular limits in
certain modulus parameters such that $S^4$ becomes either a flat
four-space $\R^4$, or else $S^3\times \R$.  In the flat space limit,
the M-theory becomes a massive supergravity in $D=7$ with a
topological massive 3-form potential.  In the $S^3\times \R$ limit, the
M-theory becomes a gauged supergravity, but with no massive 3-form
potential.  This observation give a geometrical interpretation of the
massive, but ungauged supergravity in $D=7$ and the gauged massless
$D=7$ gauged supergravity, in terms of taking limits of certain modulus
parameters.  We also obtain new domain-wall black holes
in these limiting cases.  For appropriate choice of parameters these
solutions are supersymmetric.

   We begin this section with a review of the Kaluza-Klein reduction
from $D=11$ to $D=7$ on $S^4$.  As we said in the introduction, it is
believed, although it is strictly speaking still only a conjecture,
that the maximal $N=4$, $SO(5)$ gauged supergravity in $D=7$ can be
obtained by performing a Kaluza-Klein reduction on $S^4$ accompanied
by a truncation to the massless supermultiplet of fields.  We shall
consider the further restriction discussed in \cite{ten}, where the
$SO(5)$ gauge fields are truncated down to the abelian $U(1)\times
U(1)$ subgroup, and correspondingly just two of the scalar fields of
the full $N=4$ theory are retained.  The full non-linear ansatz for
this reduction, found in \cite{ten}, is given by
\bea
ds_{11}^2 &=& \wtd\Delta^{1/3}\, ds_7^2 + g^{-2}\, \wtd\Delta^{-2/3}\,
\Big(X_0^{-1}\, d\mu_0^2 + \sum_{i=1}^2 X_i^{-1}\, (d\mu_i^2 + \mu_i^2\,
(d\psi_i + g\, A_\1^i)^2) \Big)\ ,\label{s4metred}\\
{*F_\4} &=&
2g\,\sum_{\a=0}^2 \Big(X_\a^2\, \mu_\a^2 - \wtd\Delta\, X_\a \Big)\,
\ep_\7 + g\, \wtd\Delta\, X_0\, \ep_\7
+\fft1{2g}\, \sum_{\a=0}^2 X_\a^{-1}\, {{\bar *}dX_\a}
\wedge d(\mu_\a^2) \nn\\
&&+\fft1{2g^2}\, \sum_{i=1}^2 X_i^{-2}\, d(\mu_i^2)\wedge
(d\psi_i + g\, A^i_\1) \wedge
{{\bar *} F_\2^i}\ ,\label{s4f4red}
\eea
where  ${\bar *}$ denotes the Hodge dual
with respect to the seven-dimensional metric $ds_7^2$, $\ep_\7$
denotes its volume form, and $*$ denotes the Hodge dualisation in the
eleven-dimensional metric.  The quantity $\wtd\Delta$ is given by
\be
\wtd\Delta = \sum_{\a=0}^2 X_\a\, \mu_\a^2\ ,\label{deltadef}
\ee
where $\mu_0$, $\mu_1$ and $\mu_2$ satisfy
\be
\mu_0^2+\mu_1^2+\mu_2^2=1\ .\label{muconst1}
\ee
fields are parameterised by $X_1$ and $X_2$, with $X_0$ introduced for
convenience as an auxiliary variable, defined by
$X_0\equiv (X_1 X_2)^{-2}$. The two scalar fields $X_i$ can be
parameterised in terms of two canonically-normalised dilatons
$\vec\phi=(\phi_1,\phi_2)$ by writing
\be
X_i = e^{-\fft12\vec a_i\cdot\vec\phi}\ ,\label{xidef}
\ee
where the dilaton vectors satisfy the relations $\vec a_i\cdot\vec a_j
= 4\delta_{ij} -\ft85$.  A convenient parameterisation is given by
\be
\vec a_1 = (\sqrt2, \sqrt{\ft25})\ ,\qquad \vec a_2 = (-\sqrt2,
\sqrt{\ft25}) \ .\label{d7adef}
\ee

    The $\mu_\a$ and $\psi_i$ coordinates parameterise the
compactifying 4-sphere.  When the seven-dimensional gauge fields and scalars
are set to zero, the compactifying metric becomes simply
\be
d\Omega_4^2 = d\mu_0^2 + \sum_{i=1}^2 \Big( d\mu_i^2 + \mu_i^2\,
d\psi_i^2 \Big)\ ,
\ee
which is nothing but the metric on a unit-radius round
4-sphere.

   For future reference, note that more generally, the
metric on any even-dimensional unit sphere $S^{2n}$ can be written as
\cite{mype}
\be
ds^2 = d\mu_0^2 + \sum_{i=1}^n  \Big( d\mu_i^2 + \mu_i^2\,
d\psi_i^2 \Big)\ ,\label{evensphere}
\ee
where $\sum_{\a=0}^n \mu_\a^2=1$.  For odd-dimensional spheres, the
unit $S^{2n-1}$ metric can be written as \cite{mype}
\be
ds^2 =  \sum_{i=1}^n  \Big( d\mu_i^2 + \mu_i^2\,
d\psi_i^2 \Big)\ ,\label{oddsphere}
\ee
where $\sum_{i=1}^n \mu_i^2 =1$.

    It was shown in \cite{ten} that after substituting
(\ref{s4metred}) and (\ref{s4f4red}) into the eleven-dimensional
equations of motion, one obtains seven-dimensional equations that can
be derived from the Lagrangian
\be
e^{-1}{\cal L}_7 = R -\ft12 (\del\vec\phi)^2 - V -
\ft14 \sum_{i=1}^2 e^{\vec a_i\cdot\vec\phi}\, (F_\2^i)^2\ ,
\label{d7lag}
\ee
where the potential $V$ is given by
\be
V= g^2(\, -4 X_1 X_2 - 2X_1^{-1}\, X_2^{-2} - 2 X_2^{-1}\, X_1^{-2} +\ft12
(X_1 X_2)^{-4})\ .\label{d7scalpot}
\ee
The system can be further consistently truncated to $X_1=X_2\equiv
X=e^{-\phi_2/\sqrt{10}}$.   If we make a constant shift of the
dilaton $\phi_2$ such that $X\rightarrow X\, e^{\lambda/5}$, and
define
\be
m'= g\, e^{-\ft45\lambda}\ ,\qquad g' = g\, e^{\ft15\lambda}
\ ,
\ee
the potential $V$ becomes exactly the same as the one given in
(\ref{d7gauge}), after dropping the primes on the rescaled coupling
constants.

\subsection{Limit to flat space}

     In this subsection, we consider a limiting procedure in which the
$S^4$ reduction reviewed above becomes a reduction on a flat internal
space, giving rise to a seven-dimensional theory that is equivalent to
one obtained by a certain toroidal compactification from $D=11$.    To
do this, it is useful first to apply the appropriate limiting
procedure in the seven-dimensional theory itself.

    From (\ref{xidef}) and (\ref{d7adef}), we see that the scalar
potential (\ref{d7scalpot}) is given by
\be
V = g^2\, \Big(-4 e^{-\fft12(\vec a_1 + \vec a_2)\cdot\vec\phi}
- 2e^{\fft12(\vec a_1 + 2 \vec a_2)\cdot\vec\phi}
- 2e^{\fft12(\vec 2a_1 + \vec a_2)\cdot\vec\phi} + \ft12
e^{2(\vec a_1 + \vec a_2)\cdot\vec\phi}\Big)\ .\label{d7pot2}
\ee
We now make the following redefinitions of fields and the gauge
coupling constant $g$:
\be
\vec\phi = \vec\phi' - \ft12(\vec a_1 + \vec a_2)\, \lambda\ ,\qquad
A_\1^i = e^{\fft15\lambda}\, {A_\1^i}'\ ,\qquad
g= e^{\fft45\lambda}\, m'\ ,\label{rescale1}
\ee
where $\lambda$ is a constant.
It follows from (\ref{d7adef}) and (\ref{xidef}) that $X_1$ and $X_2$,
and the auxiliary quantity $X_0$, will suffer the rescalings
\be
X_i= e^{\fft15\lambda}\, X_i'\ ,\qquad
X_0= e^{-\fft45\lambda}\, X_0'\ ,\label{rescale2}
\ee
where $i$ runs over the values 1 and 2.
Substituting into the seven-dimensional Lagrangian (\ref{d7lag}), and
then taking the limit $\lambda\longrightarrow -\infty$, we arrive at
the Lagrangian
\be
e^{-1}{\cal L}_7 = R -\ft12 (\del\vec\phi)^2 - \ft12 m^2\,
e^{\fft{8}{\sqrt{10}}\phi_2}-
\ft14 \sum_{i=1}^2 e^{\vec a_i\cdot\vec\phi}\, (F_\2^i)^2\ ,
\label{d7lag2}
\ee
where, having taken the limit, we have then suppressed the primes on
the redefined fields.  This corresponds to the limit where the
parameter $g$ in (\ref{d7gauge}) is set to zero.

    The theory described by the Lagrangian (\ref{d7lag2}) is one that
can be obtained by means of a toroidal reduction from $D=11$, in which
the usual Kaluza-Klein ansatz is generalised somewhat by allowing the
inclusion of a constant 4-volume term $q\, \ep_\4$ in the ansatz for
the eleven-dimensional $F_\4$.  In terms of an ansatz on the 3-form
potential, this corresponds to allowing a linear dependence on one or
more of the toroidal compactification coordinates \cite{lpdomain,llp}.
Here, our goal will be to show how this theory can be obtained by
taking an appropriate limit in the $S^4$ reduction described above.

   The required limiting process that we shall apply to the $S^4$
reduction ansatz given in (\ref{s4metred}) and (\ref{s4f4red}) is
governed by what we already determined in $D=7$, where the appropriate
scalings of fields and the gauge-coupling $g$ were established.  It is
evident that to obtain a regular limit, it is necessary also to apply
an appropriate rescaling to the coordinates $\mu_\a$ involved in the
parameterisation of the 4-sphere.  We find that the necessary
rescalings are as follows:
\be
\mu_0 = \mu_0'\ ,\qquad \mu_i = e^{\fft12\lambda}\, \mu_i'\ .
\label{rescale3}
\ee
Note that as we take the limit $\lambda\longrightarrow -\infty$, the
quadratic constraint (\ref{muconst1}), which becomes ${\mu_0'}^2 +
e^\lambda\, \mu_i'\, \mu_i' =1$, will imply that ${\mu_0'}^2=1$, while
the two quantities $\mu_i'$ will become unconstrained.  Also, the
quantity $\wtd\Delta$ defined in (\ref{deltadef}) will, in the limit,
be given by $\wtd\Delta \longrightarrow e^{-\fft45\lambda}\, X_0'$.

    Applying this limiting procedure to the $S^4$ reduction ans\"atze
(\ref{s4metred}) and (\ref{s4f4red}), we therefore find that in the
limit where $\lambda\longrightarrow -\infty$, the dominant terms in
the eleven-dimensional metric and 4-form become
\bea
ds_{11}^2 &=& e^{-\fft4{15}\lambda}\, \Big(
X_0^{1/3}\, ds_7^2 + m^{-2}\, X_0^{-2/3}\,
\sum_{i=1}^2 X_i^{-1}\, (d\mu_i^2 + \mu_i^2\,
d\psi_i^2)\Big) \ ,\label{s4metred2}\\
{*F_\4} &=&e^{-\fft45\lambda}\, \Big(
m\, X_0^2\, \ep_\7
+\fft1{2m^2}\, \sum_{i=1}^2 X_i^{-2}\, d(\mu_i^2)\wedge
d\psi_i \wedge
{{\bar *} F_\2^i} \Big)\ .\label{s4f4red2}
\eea
Here, we have dropped the primes on all the fields and the coupling
constant, after having taken the limit.

    Note that the overall $\lambda$-dependent prefactors in the above
expressions, although divergent as $\lambda\longrightarrow -\infty$,
do not give rise to any singularity in the seven-dimensional equations
of motion.  The reason for this is that there is a scaling symmetry of
the eleven-dimensional supergravity equations of motion, referred to
as a ``trombone'' symmetry in \cite{trombone}, under which the metric
and the 3-form potential scale by the constant factors
\be
g_{MN} \longrightarrow k^2\, g_{MN}\ ,\qquad
A_\3 \longrightarrow k^3\, A_\3\ .
\ee
(More generally, any theory with an Einstein-Hilbert term and
quadratic field-strength kinetic terms will have such a scaling
symmetry if the metric and the gauge potentials each scale with a
power of $k$ equal to the number of indices carried by the metric or
potential.)  This is a symmetry of the equations of motion.  It is not
a symmetry of the action, which scales by a uniform constant factor
under this transformation.  Bearing in mind that the ansatz for $F_\4$
given in (\ref{s4f4red2}) is for the dual of $F_\4$, which is a 7-form
in $D=11$ and thus would have a 6-form potential, the powers of
$e^\lambda$ in the prefactors in (\ref{s4metred2}) and
(\ref{s4f4red2}) are seen to be precisely of the correct form to factor out in
the equations of motion by virtue of the scaling symmetry.

   It is now quite straightforward to see that in this limiting
situation, the metric and 4-form ans\"atze are in fact equivalent to
standard ones on a 4-torus.  To see this, let us define new
coordinates $(z_1,z_2,z_3,z_4)$ in place of
$(\mu_1,\mu_2,\psi_1,\psi_2)$, where
\bea
&&z_1 = \mu_1\, \cos\psi_1\ ,\qquad z_2 = \mu_1\, \sin\psi_1\ ,\nn\\
&&z_3 = \mu_2\, \cos\psi_2\ ,\qquad z_4 = \mu_2\, \sin\psi_2\ ,
\eea
in terms of which the metric ansatz (\ref{s4metred2}) becomes
\be
ds_{11}^2 = e^{-\fft4{15}\lambda}\, \Big( X_0^{1/3}\, ds_7^2
+ m^2\, X_0^{-2/3}\, \big( X_1^{-1}\, (dz_1^2 + dz_2^2)
+ X_2^{-1}\, (dz_3^2 + dz_4^2) \big) \Big)\ .\label{s4metred3}
\ee
Similarly, we may express the 4-form ansatz in terms of the $z$
coordinates.  It is convenient at the same time to perform a
dualisation, so that we express the ansatz on the 4-form itself,
rather than its dual.  Upon doing so, we find that (\ref{s4f4red2})
becomes
\be
F_\4 = e^{-\fft25\lambda}\, \Big( m^5\, d^4z + \fft1{m^2}\, dz_3\wedge
dz_4 \wedge F_\2^1 -   \fft1{m^2}\, dz_1\wedge dz_2 \wedge F_\2^2
\Big)\ .\label{s4f4red3}
\ee

     The reduction ans\"atze (\ref{s4metred3}) and (\ref{s4f4red3})
that we have arrived at in this limiting case are equivalent to those
for a Kaluza-Klein reduction from $D=11$ on a 4-torus, with certain of
the fields set to zero, and with the inclusion of a Scherk-Schwarz
type generalisation in the reduction ansatz for the 4-form field
strength.  In such a reduction, before setting any of the
fields to zero, we would have
\bea
d\hat s_{11}^2 &=& e^{\fft13\vec g\cdot\vec\phi}\, ds_7^2 + \sum_{i=1}^4
e^{2\vec\gamma_i\cdot\vec\phi}\, (h^i)^2 \ ,\nn\\
\hat F_\4 &=& F_\4 + F_{\3 i}\, \wedge h^i +\ft12 F_{\2 ij}\wedge
h^i\wedge h^j + \ft16 F_{\1 ijk}\wedge h^i\wedge h^j\wedge h^k +
 q\, d^4z\ ,
\eea
where $h^i = dz^i + \cA_\1^i + \cA^i_{\0 j}\, dz^j$, and the various
dilaton-coupling vectors $\vec g$ and $\vec \gamma_i$ are defined in
\cite{lpsol,cjlp1}. If the Kaluza-Klein vectors $\cA_\1^i$, and their
subsequently-descendant axions $\cA^i_{\0 j}$ are truncated out, along
with certain of the other fields, it is evident that this toroidal
reduction is equivalent to the reduction defined by the ans\"atze
(\ref{s4metred3}) and (\ref{s4f4red3}).  Specifically, the two 2-form
field strengths that survive in (\ref{s4f4red3}) correspond to the two
field strengths $F_{\2 12}$ and $F_{\2 34}$ of the toroidal reduction.
The two dilatons $\vec\phi=(\phi_1,\phi_2)$, parameterised as
$X_1$ and $X_2$ in the ans\"atze (\ref{s4metred3}) and
(\ref{s4f4red3}), correspond to two specific surviving combinations of
the four dilatons $\vec\phi$ of the 4-torus reduction.

\subsection{Limit to $S^3\times \R$}

     The other limiting case that we wish to consider corresponds to
setting the parameter $m$ to zero in (\ref{d7gauge}).  In terms of the
description of the seven-dimensional theory at the beginning of this
section, this limit is achieved by making rescalings analogous to
(\ref{rescale1}) and (\ref{rescale2}), but now given by
\be
\vec\phi = \vec\phi' - \ft12(\vec a_1 + \vec a_2)\, \lambda\ ,\qquad
A_\1^i = e^{\fft15\lambda}\, {A_\1^i}'\ ,\qquad
g= e^{-\fft15\lambda}\, g'\ ,\label{rescale4}
\ee
where $\lambda$ is a constant.  Thus the crucial change is that $g$ is
now rescaled differently from before.
It follows from (\ref{d7adef}) and (\ref{xidef}) that $X_1$ and $X_2$,
and the auxiliary quantity $X_0$, will rescale as
\be
X_i= e^{\fft15\lambda}\, X_i'\ ,\qquad
X_0= e^{-\fft45\lambda}\, X_0'\ ,\label{rescale5}
\ee
Now, if we take the limit where $\lambda\longrightarrow +\infty$, we
see that the Lagrangian (\ref{d7lag}) reduces to
\be
e^{-1}{\cal L}_7 = R -\ft12 (\del\vec\phi)^2 + 4 g^2\,
e^{- \fft{2}{\sqrt{10}}\phi_2}-
\ft14 \sum_{i=1}^2 e^{\vec a_i\cdot\vec\phi}\, (F_\2^i)^2\ ,
\label{d7lag3}
\ee
where, having taken the limit, we have again suppressed the primes on
the redefined fields.  This corresponds to the limit where the
parameter $m$ in (\ref{d7gauge}) is set to zero.

    We may again now consider the limiting procedure from the
viewpoint of the ans\"atze for the eleven-dimensional fields.  It is
easily seen in this case that we should rescale the coordinates
$\mu_\a$ in the 4-sphere as follows:
\be
\mu_0 = e^{-\fft12\lambda}\, \mu_0'\ ,\qquad \mu_i = \mu_i'\ .
\ee
This has the effect that the constraint (\ref{muconst1}) becomes
\be
e^{-\lambda}\, {\mu_0'}^2 + \mu_i'\, \mu_i'=1\ ,
\ee
where as usual, $i$ runs over 1 and 2.  In the limit when we send
$\lambda\longrightarrow+\infty$, we will therefore have
\be
\mu_i'\, \mu_i' = 1\ ,\label{mu1mu2const}
\ee
with $\mu_0'$ unconstrained.  The quantity $\wtd\Delta$ defined in
(\ref{deltadef}) will, in the limit where $\lambda\longrightarrow
+\infty$, tend to the form
\be
\wtd\Delta \longrightarrow e^{\fft15\lambda}\, \wtd\Delta'\ ,
\ee
where
\be
\wtd\Delta' = \sum_{i=1}^2 X_i'\, {\mu_i'}^2\ .
\ee

    Applying this limiting procedure to the ans\"atze (\ref{s4metred})
and (\ref{s4f4red}) for the eleven-dimensional fields, we find that
they become
\bea
ds_{11}^2 &=& e^{\fft1{15}\lambda}\Big\{
\wtd\Delta^{1/3}\, ds_7^2 \nn\\
&&+ g^{-2}\, \wtd\Delta^{-2/3}\,
\Big(X_0^{-1}\, d\mu_0^2 + \sum_{i=1}^2 X_i^{-1}\, (d\mu_i^2 + \mu_i^2\,
(d\psi_i + g\, A_\1^i)^2) \Big)\Big\}\ ,\label{s4metred4}\\
{*F_\4} &=&e^{\ft15\lambda}\Big\{
2g\,\sum_{i=1}^2 \Big(X_i^2\, \mu_i^2 - \wtd\Delta\, X_i \Big)\,
\ep_\7
+\fft1{2g}\, \sum_{i=1}^2 X_i^{-1}\, {{\bar *}dX_\a}
\wedge d(\mu_i^2) \nn\\
&&+\fft1{2g^2}\, \sum_{i=1}^2 X_i^{-2}\, d(\mu_i^2)\wedge
(d\psi_i + g\, A^i_\1) \wedge
{{\bar *} F_\2^i} \Big\}\ ,\label{s4f4red4}
\eea
where again we have suppressed the primes on the various fields
and coordinates, and on $g$, after having taken the limit.

    The four coordinates $\mu_i$ and $\psi_i$, together with the
quadratic constraint (\ref{mu1mu2const}), can be recognised as
parameterising a 3-sphere in the internal space
(see equation (\ref{oddsphere})).  The coordinate
$\mu_0$, which is now unconstrained, parameterises the fourth of the
directions in the internal space.  Initially, the $\mu_0$ coordinate
was one of three that were subject to the quadratic constraint
(\ref{muconst1}) in the original 4-sphere.  A parameterisation of the
original $\mu_\a$ in terms of two unconstrained angles could be taken
to be $\mu_0=\cos\theta_1$, $\mu_1=\sin\theta_1\, \cos\theta_2$,
$\mu_2 = \sin\theta_1\, \sin\theta_2$.  Thus
$\mu_0$ originally ranged over a finite line segment.  As the limit
$\lambda\longrightarrow +\infty$ is taken, this line segment expands
to cover the entire real line.

    We can use the $S^3\times \R$ limit of the $S^4$ reduction ansatz
that we have just obtained in order to describe an $S^3$ reduction of
ten-dimensional type IIA supergravity.  To do this, we break the
$S^3\times \R$ reduction up into two steps, the first of which
consists of reducing from $D=11$ to $D=10$ on the $\R$ direction.  Of
course since this corresponds to a Killing symmetry of the ansatz,
generated by $\del/\del\mu_0$, we can choose to re-interpret $\mu_0$
as an angular ignorable coordinate on $S^1$, rather than a coordinate
on the real line.  At the same time, we can exploit the trombone
scaling symmetry of the $D=11$ equations of motion to eliminate the $
e^{\lambda}$ factors in the limiting form of the ans\"atze.

    To implement the resulting $S^1$ reduction, we use the standard
Kaluza-Klein metric ansatz
\be
d\hat s_{11}^2 = e^{-\fft16\phi}\, ds_{10}^2 + e^{\fft43\phi}\, (dz +
\cA_\1)^2\ .
\ee
For the 4-form, we have $\hat F_4 = F_\4 + F_\3\wedge (dz+\cA_\1)$,
with $F_4 = dA_\3 - dA_\2\wedge \cA_\1$ and $F_\3=dA_\2$.  From these
ans\"atze, it is easy to see that we shall have
\be
{\hat * \hat F_\4} = e^{\fft12\phi}\, {*F_\4}\wedge (dz+\cA_\1)  +
e^{-\phi}\, {* F_\3}\ ,
\ee
where we are using $\hat *$ here to denote an eleven-dimensional Hodge
dual, and $*$ to denote a ten-dimensional one.

    From the above, it follows that the $S^3\times \R$ metric and
4-form reduction ans\"atze (\ref{s4metred4}) and (\ref{s4f4red4})
can be re-interpreted as type IIA $S^3$ reduction ans\"atze with
\bea
&&ds^2_{10} = \wtd \Delta^{1/4}\, (X_1 X_2)^{1/4}\, ds_7^2 +
g^{-2} \wtd \Delta^{-3/4}\, (X_1 X_2)^{1/4}\,
\sum_{i=1}^2 X_i^{-1}\, (d\mu_i^2 + \mu_i^2\, (d\psi_i + g\,
A_\1^i)^2)\ ,\nn\\
&&e^{-\phi}\, {*F_\3}=
2g\,\sum_{i=1}^2 \Big(X_i^2\, \mu_i^2 - \wtd\Delta\, X_i \Big)\,
\ep_\7
+\fft1{2g}\, \sum_{i=1}^2 X_i^{-1}\, {{\bar *}dX_i}
\wedge d(\mu_i^2) \nn\\
&&\qquad+\fft1{2g^2}\, \sum_{i=1}^2 X_i^{-2}\, d(\mu_i^2)\wedge
(d\psi_i + g\, A^i_\1) \wedge
{{\bar *} F_\2^i} \ , \label{s3d10red}\\
&&e^{2\phi} = \fft{(X_1X_2)^3}{\wtd \Delta}\ ,\qquad F_\4=0\ ,\qquad
\cF_\2=0\ ,\nn
\eea
where $\phi$ is the dilaton of the type IIA theory.

    We arrived at the above ans\"atze for the $S^3$ reduction of type
IIA supergravity by taking a singular limit of the $S^4$ reduction of
$D=11$ supergravity.  However, having obtained it by this method, we can
now view it as a valid reduction procedure in its own right.  In
particular, by virtue of the previously-established consistency of the
$S^4$ reduction, we now know that this procedure has provided us with
a consistent $S^3$ reduction scheme for the type IIA theory, leading
to the seven-dimensional Lagrangian given in (\ref{d7lag3}).
In particular, this is an
example of a non-trivial sphere reduction (\ie one that includes
scalars describing inhomogeneous sphere deformations) that gives rise
to a gauged supergravity with no AdS vacuum.

    Note that the Ramond-Ramond fields $F_\4$ and $\cF_\2$ are set to
zero in the ans\"atze (\ref{s3d10red}) for the $S^3$ reduction of type
IIA supergravity.  This means that we can also interpret it as an
$S^3$ reduction of the $N=1$ or heterotic theories in $D=10$.
Although $\mu_0$ naturally became a coordinate on the entire real line
in the taking of the singular limit of the $S^4$ compactification, the
consistency of the Kaluza-Klein reduction does not of itself require
that $\mu_0$ cover the infinite interval.  We could instead impose a
cut-off on the interval, and require instead that $\mu_0$ range only
over a finite line segment, $S^1/Z_2$.  This could then be combined
with a projection in which the Ramond-Ramond fields were set to zero,
precisely in the fashion of the $S^1/Z_2$ reduction of M-theory
introduced in \cite{horava}.  Thus the reduction ans\"atze
(\ref{s4metred4}) and (\ref{s4f4red4}), where the Ramond-Ramond fields
in $D=10$ have already been projected out, can also be viewed as a
consistent reduction ansatz for M-theory on $S^3\times (S^1/Z_2)$.
The limit that was taken in arriving at (\ref{s4metred4}) and
(\ref{s4f4red4}) from the $S^4$ reduction ansatz however implies that
the line segment $S^1/Z_2$ expands to become $\R$; this corresponds to
strongly coupled heterotic string theory (and the topological information
encoded in $S^1/Z_2$ is lost).

\subsection{New domain-wall black holes in $D=7$}

    Having obtained the two singular limits of the 4-sphere
reduction of eleven-dimensional supergravity, it is of interest to
study how the solutions behave in these limits.  One example was
already illustrated in the introduction, where a domain-wall solution
supported purely by the scalar potential was presented.  As we
discussed there, the limits associated with $m=0$ and $g=0$ give
rise to domain walls that can be respectively interpreted as the near
horizon of an isotropic NS-NS 5-brane or an M5-brane, with charges
uniformly distributed over a flat four-space.  In this subsection, we
obtain a new domain-wall black-hole solution, which is not merely
the singular-modulus limit of the previously known AdS black hole.

           The Lagrangian (\ref{d7gauge}) admits AdS$_7$ black
hole solutions, given by
\bea
ds_7^2&=& -H^{-8/5}\, f\, dt^2 + H^{2/5}\, (f^{-1} dr^2 + r^2\,
d\Omega_{5,k}^2)\ ,\nn\\
f&=& k -\fft{\mu}{r^4} + \ft14 m^2 r^2\, H^2\ ,\qquad
e^{\fft{5}{\sqrt{10}} \phi} = H\ ,\nn\\
A_\1 &=&{\sqrt{2}(1+k\sinh^2\alpha)^{1/2}\over\sinh\a}\, H^{-1}\, dt
\ ,\qquad H= \fft{g}{m}\left(1 + \fft{\mu\, \sinh^2\a}{r^4}\right)\ ,
\label{d7adsbh}
\eea
where $d\Omega^2_{5,k}$ is the metric on a unit $S^5$, $T^5$ or $H^5$
according to whether $k=1,0$ or $-1$.  (Note that the solution was
previously obtained in \cite{ten,lm} for $g=m$.) In particular the
$k=0$ solution can be oxidised back to $D=11$ and becomes
\cite{Cveticgubser1,ten} the near horizon structure of the rotating
M5-brane \cite{Cveticyoum2}.  We see that the solution (\ref{d7adsbh})
does not admit an $m=0$ limit.  It does, however, allow us to take the
limit $g\to0$, provided that $g\sinh^2\alpha\equiv Q$ is held
fixed.  The solution then reduces to a domain-wall black hole.

      Although the solution (\ref{d7adsbh}) does not admit an $m=0$
limit, there does exist a new domain-wall black hole solution for the
Lagrangian (\ref{d7gauge}) when $m=0$.  In fact when $m=0$, the
Lagrangian fits precisely the general pattern in
(\ref{gensinglelagsp}) in the appendix.  It follows that the
domain-wall black hole is given by
\bea
\label{eq:7ddwbh}
ds_7^2&=& -f\, dt^2 + f^{-1}\, dr^2 + r\, (dx_1^2 + \cdots + dx_5^2)
\ ,\nn\\
f&=& 2r(\ft8{25} g^2 + \fft{\mu}{r^{5/2}} + \fft{\lambda^2}{4r^5})
\ ,\qquad e^{\ft{2}{\sqrt{10}}\,\phi} = r\ ,\nn\\
A_\1 &=& \lambda\, r^{-5/2}\, dt\ .
\eea

To examine the supersymmetry of this domain-wall black hole, we note that
the $N=2$ supersymmetry transformations on the fermions for the Lagrangian
(\ref{d7gauge}) are given by \cite{lm}
\begin{eqnarray}
\delta\psi_\mu &=& [\nabla_\mu+\ft{\im}{\sqrt{2}}gA^a_\mu\sigma^a
+\ft1{20}(me^{\fft{4}{\sqrt{10}}\phi} +4ge^{-\fft1{\sqrt{10}}\phi})\gamma_\mu
-\ft{\im}{20\sqrt{2}}(\gamma_\mu{}^{\nu\lambda}-8\delta_\mu^\nu\gamma^\lambda)
e^{\fft{1}{\sqrt{10}}\phi}F^a_{\nu\lambda}\sigma^a\nonumber\\
&&\qquad\qquad\qquad+\ft1{160}(\gamma_\mu{}^{\nu\rho\sigma\lambda}
-\ft83\delta_\mu^\nu\gamma^{\rho\sigma\lambda})
e^{-\fft{2}{\sqrt{10}}\phi} F_{\nu\rho\sigma\lambda}
]\epsilon\ ,\nonumber\\
\delta\lambda&=&[-\ft1{2\sqrt{2}}\gamma^\mu\partial_\mu\phi
+\ft1{\sqrt{5}}(m\, e^{\fft{4}{\sqrt{10}}\phi}
-g\, e^{-\fft{1}{\sqrt{10}}\phi})
+\ft{\im}{4\sqrt{10}}e^{\fft{1}{\sqrt{10}}\phi}
F^a_{\mu\nu}\gamma^{\mu\nu}\sigma^a\nonumber\\
&&\qquad\qquad\qquad
+\ft1{48\sqrt{5}} e^{-\fft{2}{\sqrt{10}}\phi}F_{\mu\nu\rho\sigma}
\gamma^{\mu\nu\rho\sigma}]\epsilon\ ,
\end{eqnarray}
where the $\sigma^a$ are the $SU(2)$ Pauli matrices.  In the absence
of $F_{(4)}$, and with only the $U(1)$ subgroup of $SU(2)$ excited (as
is appropriate for the above black hole solutions), the supersymmetry
transformations reduce to
\begin{eqnarray}
\delta\psi_\mu &=& [\nabla_\mu+\ft{\im}{\sqrt{2}}gA_\mu
+\ft1{20}(me^{\fft{4}{\sqrt{10}}\phi} +4ge^{-\fft1{\sqrt{10}}\phi})\gamma_\mu
-\ft{\im}{20\sqrt{2}}(\gamma_\mu{}^{\nu\lambda}-8\delta_\mu^\nu\gamma^\lambda)
e^{\fft{1}{\sqrt{10}}\phi}F_{\nu\lambda}]\epsilon\ ,\nonumber\\
\delta\lambda&=&[-\ft1{2\sqrt{2}}\gamma^\mu\partial_\mu\phi
+\ft1{\sqrt{5}}(m\, e^{\fft{4}{\sqrt{10}}\phi}
-g\, e^{-\fft{1}{\sqrt{10}}\phi})
+\ft{\im}{4\sqrt{10}}e^{\fft{1}{\sqrt{10}}\phi}
F_{\mu\nu}\gamma^{\mu\nu}]\epsilon\ .
\end{eqnarray}
(Here, we have adopted the natural complex notation for the
supersymmetry parameter, in this $U(1)$ truncation.)

   Examination of the $\delta\lambda$ variation
indicates that the domain-wall black hole (\ref{eq:7ddwbh}) is in general
non-supersymmetric unless $\mu=0$, so that
$f=\fft{16}{25}g^2r+\fft12\lambda^2r^{-4}$.  In this case the Killing
spinors must satisfy the half-supersymmetry projection $P\epsilon = 0$ where
\begin{equation}
P=\ft12[1+f^{-1/2}(\ft45gr^{1/2}\gamma_1-\ft{\im}{\sqrt{2}}\lambda r^{-2}
\gamma_0)]\ .
\end{equation}
Note that $\gamma_0$ and $\gamma_1$ denote the Dirac matrices with vielbein
indices in the $t$ and $r$ directions respectively.
For such spinors $\epsilon$, the gravitino variation reduces to
\begin{eqnarray}
\delta\psi_0&=&\partial_0\epsilon\ ,\nonumber\\
\delta\psi_r&=&[\partial_r+\fft{1}{r}+gf^{-1/2}r^{-1/2}\gamma_1]\epsilon\ ,
\nonumber\\
\delta\psi_{\theta_i}&=&\partial_{\theta_i}\epsilon\ .
\end{eqnarray}
The above equations are easily solved to obtain the $N=2$ Killing spinors
\begin{equation}
\label{eq:7dkil}
\epsilon=[\sqrt{f^{1/2}-\ft1{\sqrt{2}}\lambda r^{-2}}-
\sqrt{f^{1/2}+\ft1{\sqrt{2}}\lambda r^{-2}}\,\gamma_1](1+\im\gamma_0)
\epsilon_0 \ ,
\end{equation}
where $\epsilon_0$ is an arbitrary constant spinor.  Note that the
projection $P_0\equiv\fft12(1+\im\gamma_0)$ indicates that the
domain-wall black hole preserves exactly half of the supersymmetries
when $\mu=0$.

     To summarise, we have seen that there are two distinct domain-wall
black hole solutions in $D=7$, corresponding to the limiting cases where
$g=0$ or $m=0$.  The former, describing a domain-wall black hole with $k=1$,
corresponds to a solution of the theory obtained by compactifying $D=11$
supergravity on $\R^4$; this itself, as we saw, is a singular limit of
the $S^4$ compactification.  The other domain-wall black hole, with $m=0$,
has $k=0$.  This corresponds to a solution of the $S^3\times \R$
compactification from $D=11$, which is another singular limit of the $S^4$
compactification.

\section{$S^7$ reduction of M-theory, and its limits}

   Let us now turn to the four-dimensional theory obtained by the
dimensional reduction of M-theory on $S^7$.  We shall consider two limits
that correspond to consistent reductions on $S^3\times\R^4$ and
$S^5\times\R^2$.  We also find new domain-wall black hole solutions.

As shown in \cite{ten},
there is an $N=2$, $U(1)^4$ truncation of the full $N=8$ gauged
supergravity, for which the explicit Kaluza-Klein reduction ansatz can
be given.\footnote{A fully consistent truncation would require the
inclusion of 3 axionic scalars as well as the 3 dilatonic scalars that
are present in the $U(1)^4$ ansatz given in \cite{ten}.  If one is
considering solutions such as the 4-charge AdS black holes of $D=4$
gauge supergravity \cite{Duffliu}, for which the axions can be set to
zero, this truncation of the axions is justified.}  The reduction
ansatz is
\bea
ds_{11}^2 &=& \wtd\Delta^{2/3}\, ds_4^2 +g^{-2}\,
\wtd\Delta^{-1/3}\, \sum_{i=1}^4 X_i^{-1}\, \Big( d\mu_i^2 + \mu_i^2\,
 (d\psi_i + g\,
A^i_\1)^2 \Big)\ ,\label{s7metred}\\
F_\4 &=& 2g\,\sum_{i=1}^4 \Big(X_i^2\, \mu_i^2 - \wtd\Delta\, X_i \Big)\,
\ep_\4 +\fft1{2g}\, \sum_{i=1}^4 X_i^{-1}\, {{\bar *}dX_i}\wedge d(\mu_i^2)
\nn\\
&&-\fft1{2g^2}\, \sum_{i=1}^4 X_i^{-2}\, d(\mu_i^2)\wedge (d\psi_i + g\,
A^i_\1) \wedge
{{\bar *} F_\2^i}\ ,\label{s7f4red}
\eea
where $\wtd \Delta = \sum_{i=1}^4 X_i\, \mu_i^2$.  The four quantities
$\mu_i$ satisfy $\sum_i \mu_i^2 =1$.  Here, ${\bar *}$ denotes the
Hodge dual with respect to the four-dimensional metric $ds_4^2$, and
$\ep_\4$ denotes its volume form.

    The four $X_i$, which satisfy $X_1X_2X_3X_4=1$, can be parameterised
in terms of three dilatonic scalars $\vec\phi =(\phi_1,
\phi_2, \phi_3)$ as
\be
X_i=e^{-\ft12 \vec a_i \cdot \vec \phi}\ ,
\ee
where the $\vec a_i$ satisfy the dot products
\be
M_{ij} \equiv \vec a_i\cdot \vec a_j = 4\delta_{ij} -1\ .
\ee
A convenient choice is
\be
\vec a_1 =(1,1,1)\ ,\quad \vec a_2=(1,-1,-1)\ ,\quad
\vec a_3=(-1,1,-1)\ ,\quad \vec a_4=(-1,-1,1)\ .
\ee
In terms of this basis, the scalar potential in $D=4$ is given by
\be
V = - 4g^2\, \sum_{i<j} X_i\, X_j = -8 g^2\, (\cosh\phi_1 +
\cosh\phi_2 + \cosh\phi_3)\ ,
\ee
and in fact the reduction ansatz given above leads \cite{ten} to the
four-dimensional theory described by the Lagrangian
\be
e^{-1}{\cal L}_4 = R -
\ft12 (\del\vec\phi)^2 +
8g^2 (\cosh\phi_1 +\cosh\phi_2+\cosh\phi_3)
-\ft14 \sum_{i=1}^4 e^{\vec a_i\cdot\vec \phi}\,
(F_\2^i)^2\ .\label{d4lagxx}
\ee

\subsection{Limit to $S^3\times \R^4$}

     Let us first consider a limit where the dilaton $\phi_1$ is
subjected to the following shift:
\be
\phi_1 = \phi_1' - \lambda\ ,
\ee
with $\phi_2$ and $\phi_3$ unchanged.  This will give a finite
Lagrangian as $\lambda\longrightarrow +\infty$, provided that the
coupling constant and the gauge potentials are scaled as follows:
\be
g = e^{-\fft12\lambda}\, g'\ , \qquad A_\1^a = e^{\fft12\lambda}\,
{A_\1^a}'\ ,\qquad A_\1^{\bar a} = e^{-\fft12\lambda}\, {A_\1^{\bar a}}'
\ ,
\ee
where we have introduced the notation that the index $i=(1,2,3,4)$ is
split into $a=(1,2)$ and $\bar a=(3,4)$.  Note that the quantities $X_i$
scale as
\be
X_a = e^{\fft12\lambda}\, X_a'\ ,\qquad
X_{\bar a} = e^{-\fft12\lambda}\, X_{\bar a}'\ .
\ee
After these redefinitions, the scalar potential in
(\ref{d4lagxx}) becomes simply
\be
V = -4 {g'}^2\, e^{-\phi_1'}\ .\label{d4trunc1}
\ee

    We find that these scalings can be implemented in the metric and
4-form ans\"atze (\ref{s7metred}) and (\ref{s7f4red}) if we also make
the scalings
\be
\mu_a = \mu_a'\ ,\qquad \mu_{\bar a} = e^{-\fft12\lambda}\, \mu_{\bar
a}'\ .
\ee
In particular, this means that the condition $\sum_i \mu_i^2=1$
becomes, in the limit $\lambda\longrightarrow +\infty$, simply
$\sum_{a} {\mu_{a}'}^2=1$, with $\mu_{\bar a}'$ unconstrained.
Defining $\wtd\Delta' = \sum_{a} X_{a}'\, {\mu_{
a}'}^2$, we see that as $\lambda\longrightarrow +\infty$, the
ans\"atze become
\bea
ds_{11}^2 &=& e^{\fft13\lambda}\,
\Big\{\wtd\Delta^{2/3}\, ds_4^2 +g^{-2}\,
\wtd\Delta^{-1/3}\,\Big( \sum_{a=1}^2 X_a^{-1}\,\big( d\mu_a^2 + \mu_a^2\,
 (d\psi_a + g\, A_\1^a)^2 \big)\nn\\
&&\qquad +  \sum_{\bar a=3}^4 X_{\bar a}^{-1}\,( d\mu_{\bar a}^2
+ \mu_{\bar a}^2\, d\psi_{\bar a}^2 \Big) \Big\}
\ ,\label{s7metred2}\\
F_\4 &=& e^{\fft12\lambda}\, \Big\{
2g\,\sum_{a=1}^2 \Big(X_{a}^2\, \mu_{a}^2 -
\wtd\Delta\, X_{a} \Big)\,
\ep_\4 +\fft1{2g}\, \sum_{a=1}^2 X_{a}^{-1}\,
{{\bar *}dX_{a}}\wedge d(\mu_{a}^2)
\nn\\
&&\qquad  -\fft1{2g^2}\, \sum_{a=1}^2 X_{a}^{-2}\,
d(\mu_{a}^2)\wedge (d\psi_{a}+ g\, A_\1^{a}) \wedge
{{\bar *} F_\2^{a}} \nn\\
&&\qquad-\fft1{2g^2}\, \sum_{\bar a=3}^4
X_{\bar a}^{-2}\, d(\mu_{\bar a}^2)\wedge d\psi_{\bar a} \wedge
{{\bar *} F_\2^{\bar a}} \Big\} \ ,\label{s7f4red2}
\eea
where, as usual, we have dropped the primes after taking the limit.
This can be recognised as a reduction on $S^3\times \R^4$.

    Substituting these ans\"atze into the eleven-dimensional equations of
motion, we will obtain equations of motion that can be derived from the
following four-dimensional Lagrangian:
\begin{eqnarray}
e^{-1}{\cal L}_4 &=& R - \ft12 (\del\vec\phi)^2 + 4g^2\, e^{-\phi_1}
- \ft14 e^{\phi_1}\,
\Big( e^{\phi_2+\phi_3}\, (F_\2^1)^2 + e^{-\phi_2-\phi_3}\, (F_\2^2)^2
\Big) \nonumber\\
&&\qquad
-\ft14 e^{-\phi_1}\,
 \Big( e^{\phi_2-\phi_3}\, (F_\2^3)^2 + e^{-\phi_2+\phi_3}\,
(F_\2^4)^2 \Big)\ .
\end{eqnarray}
For simplicity, we may now consistently set $\phi_2=\phi_3=0$, provided that
we also set $F_\2^1=F_\2^2\equiv F_\2/\sqrt2$ and $F_\2^3 = F_\2^4=
{\cal F}_\2/\sqrt2$.  Thus we arrive at the truncated Lagrangian
\be
e^{-1}{\cal L}_4 = R - \ft12 (\del\phi_1)^2 -\ft14 e^{\phi_1}\,
(F_\2)^2 -\ft14 e^{-\phi_1}\, (\cF_\2)^2
+ 4 g^2\, e^{-\phi_1}\ .\label{d4lagxxxx}
\ee
The field strengths $F_\2$ and $\cF_\2$ play quite different r\^oles in
this truncated theory.  As can be seen from the ans\"atze (\ref{s7metred2})
and (\ref{s7f4red2}), the field $F_\2$, which is associated with $F_\2^a$,
is in the Cartan subgroup of the original gauged supergravity
Yang-Mills group.  On the other hand $\cF_\2$, associated with
$F_\2^{\bar a}$, is in the ungauged sector corresponding to
the $\R^4$ part of the $S^3\times \R^4$ limiting compactification.

\subsection{Domain-wall black holes in $D=4$}

         In the previous section, we singled out the scalar $\phi_1$,
making a rescaling under which the 7-sphere becomes $S^3\times R^4$.
We then consistently set $\phi_2$ and $\phi_3$ to zero,
provided that $F_\2^1=F_\2^2\equiv F_\2/\sqrt2$ and $F_\2^3 = F_\2^4=
{\cal F}_\2/\sqrt2$.  For further simplicity, we shall set ${\cal
F}_\2=0$.  In this section, we shall accordingly look for solutions to
the Lagrangian
\be
e^{-1}{\cal L}_4 = R - \ft12 (\del\phi)^2 -\ft14 e^{\phi}\,
(F_\2)^2 + 4m^2\,
e^{\phi} + 4 g^2\, e^{-\phi} + 16 m\, g\ .\label{d4lagxxx}
\ee
Note that we have introduced an additional constant parameter $m$, by
making a constant shift of the dilaton $\phi$, accompanied by a
redefinition of $g$.  For non-vanishing $m$ and $g$, the scalar
potential is equivalent to the one obtained by setting
$\phi_2=\phi_3=0$ in (\ref{d4lagxx}).   The
$N=2$ supersymmetry transformation rules for the fermionic
superpartners are given, in a purely bosonic background, by
\bea
\delta \psi^i_\mu &=& D_\mu\, \ep^i -\ft1{\sqrt2}\, g\, A_\mu\,
\ep^{ij}\, \ep_j  + \ft12 ( m\, e^{\fft12\phi} + g\, e^{-\fft12
\phi})\, \gamma_\mu\, \ep^i + \fft1{8\sqrt2}\, e^{\fft12\phi}\,
F_{\nu\rho}\, \gamma^{\nu\rho}\, \gamma_\mu\, \ep^{ij}\, \ep_j\ ,
\nn\\
\delta \chi^i &=& -\gamma^\mu\, \del_\mu\phi\, \ep^{ij}\, \ep_j +
2(m\,  e^{\fft12\phi} - g\, e^{-\fft12 \phi})\, \ep^{ij}\, \ep_j
+ \ft1{2\sqrt2}\, e^{\fft12\phi}\, F_{\mu\nu}\, \gamma^{\mu\nu}\,
\ep^i\ ,\label{d4susy}
\eea
where $\ep^i$ denotes the two supersymmetry parameters.  These can be
rewritten in terms of complex spinors as
\bea
\delta \psi_\mu &=& D_\mu\, \ep -\ft\im{\sqrt2}\, g\, A_\mu\,
\ep  + \ft12 ( m\, e^{\fft12\phi} + g\, e^{-\fft12
\phi})\, \gamma_\mu\, \ep + \fft\im{8\sqrt2}\, e^{\fft12\phi}\,
F_{\nu\rho}\, \gamma^{\nu\rho}\, \gamma_\mu\,  \ep\ ,
\nn\\
\delta \chi &=& -\im\, \gamma^\mu\, \del_\mu\phi \, \ep +
2\im\, (m\,  e^{\fft12\phi} - g\, e^{-\fft12 \phi})\, \ep
+ \ft1{2\sqrt2}\, e^{\fft12\phi}\, F_{\mu\nu}\, \gamma^{\mu\nu}\,
\ep\ ,\label{d4susy2}
\eea

      The full Lagrangian (\ref{d4lagxxx}) admits an AdS$_4$ black
hole solution, given by
\bea
\label{eq:ads4bh}
ds_4^2 &=& -H^{-1}\, f\, dt^2 + H\, (f^{-1} \, dr^2 + r^2\,
d\Omega_{2,k}^2)\ ,\nn\\
f &=& k -\fft{\mu}{r} + 4m^2r^2\, H^2\ ,\qquad e^{\phi}=H\ ,\nn\\
A_\1 &=&\fft{\sqrt{2}(1+k\sinh^2\alpha)^{1/2}}{\sinh\alpha}\, H^{-1}\, dt
\ ,\qquad
H = \fft{g}{m}\left(1 + \fft{\mu\sinh^2\a}{r}\right)\ .
\eea
Here $d\Omega_{2,k}^2$ denotes the metric on a two-dimensional space
of positive, zero or negative constant curvature, according to whether
$k$ is positive, zero or negative.  By convention, we take the spaces
with $k=+1$ and $k=-1$ to be the unit $S^2$ and hyperbolic space $H^2$
respectively.  Note that the solution with $k=1$ corresponds to a
special case of the 4-charge AdS black holes found in \cite{Duffliu},
where just two equal charges are turned on.  The solution with $k=0$
can be obtained from the $S^7$ reduction of the near horizon structure
of the rotating M2-brane \cite{ten}.  It is straightforward to
check that in the extremal limit, where $\mu$ is sent to zero while
holding $\mu\, \sinh^2\a$ fixed, these AdS black-hole solutions
preserve one half of the $N=2$ supersymmetry.

    We see that these solutions do not allow an $m=0$ limit to be
taken, but they do permit us instead to take the $g\to0$ limit provided
$\alpha\to\infty$ with $g\sinh^2\alpha\equiv Q$ held fixed.  In this limit
we find
\begin{equation}
A_\1\to\sqrt{2k}H^{-1}dt\ ,\qquad H\to{Q\mu\over m\,r}\ ,
\end{equation}
which describes an electrically charged domain-wall black hole for
non-vanishing $k$.  On the other hand, when $k=0$ the above solution is a
pure domain-wall black hole with vanishing electric charge.

    The scalar potential in (\ref{d4lagxxx}) has a symmetry
$\phi\leftrightarrow -\phi + 2\log g-2\log m$.  Accordingly we find a
solution where $e^{-\phi} =H$ rather than $e^{\phi} =H$ given by
\bea
ds_4^2 &=& - H^{-1}\, f\, dt^2 + H\, (f^{-1}\, dr^2 + r^2 \,
d\Omega_{2,k}^2) \ ,\nn\\
f &=& k -\fft{\mu}{r} + 4g^2 r^2 \, H^2 + \fft{\lambda^2}{2r^2}\ ,\qquad
e^{-\phi} = H\ ,\nn\\
A_\1&=& \fft{\lambda}{r}\, dt\ ,\qquad H = \fft{m}{g} + \fft{Q}{r}\ ,
\eea
where the constant parameters satisfy
\be
m^2\, \lambda^2+ 2 g\,  Q \, (m\, \mu + g\, k\, Q) = 0\ .\label{lamqrel}
\ee
Thus we have two possibilities.  If $m\ne0$ this solution is in fact
equivalent to (\ref{eq:ads4bh}).  This may be seen by performing
a coordinate transformation
\begin{equation}
t=\fft{m}{g}\tilde t\ ,\qquad
r=\fft{g}{m}(\tilde r-Q)\ ,
\end{equation}
and making the identification
\begin{equation}
\mu=\fft{g\, \td\mu}{m}(1+2k\sinh^2\alpha)\ ,\qquad
Q=-\tilde\mu\sinh^2\alpha\ ,\qquad
\lambda=\fft{\sqrt 2 g\tilde\mu}{m}\sinh\alpha(1+k\sinh^2\alpha)^{1/2}\ .
\end{equation}
After dropping the tilde this may be seen to be identical to
(\ref{eq:ads4bh}).  On the other hand, if $m=0$ then $\lambda$ is no longer
constrained by (\ref{lamqrel}).  In this case the constraint requires
$k=0$ for non-vanishing $Q$.  For $m=0$, $k=0$ the solution cannot be
obtained from the original AdS$_4$ black hole of (\ref{eq:ads4bh}).
Nevertheless, the Lagrangian
(\ref{d4lagxxx}) fits the pattern of the generic class of Lagrangians
(\ref{gensinglelagsp}) given in the appendix, and the solution is a
special case of (\ref{gensolsp}).  In this case, the parameter $Q$ can
be absorbed into a rescaling of coordinates.
The solution with $m=0$ was obtained in \cite{klemm}.

   From the transformation rules (\ref{d4susy2}), we find that these
$\phi=-\log H$ domain-wall black holes preserve half the $N=2$
supersymmetry if the following condition hold:
\be
Q = -\fft{m\, \mu}{2 k\, g}\ ,\qquad \lambda = \fft{\mu}{\sqrt{2k}}\ .
\label{d4susyrel}
\ee
The conditions (\ref{d4susyrel})
imply that the function $f$ in the metric is given by
\be
f= \fft{g^2}{m^2}\, (k + 4 m^2\, r^2)\, H^2\ .
\ee
When $k=1$, so that we have $S^2$ surfaces with
$d\Omega_{2,1}^2=d\theta^2+\sin^2\theta \,d\varphi^2$, the Killing spinors
take the form
\begin{equation}
\epsilon=H^{1/4}e^{-\im gt}[\sqrt{h^{1/2}+1}+\sqrt{h^{1/2}-1}\,\gamma_1]
e^{-\fft{\im}{2}\theta\gamma_{012}}e^{\fft12\varphi\gamma_{23}}
(1+\im\gamma_0)\epsilon_0\ ,
\end{equation}
where $h=1+4m^2r^2$.

    In the special case where $m=0$ and $k=0$, for which the 2-metric
$d\Omega^2_{2,0}$ can be taken to be $dy^i\, dy^i$, the solution is
supersymmetric if
\be
\mu=0\ ,\qquad H= \fft{Q}{r}\ ,\qquad f = \fft{\lambda^2}{2 r^2} + 4
g^2\, Q^2\ .
\ee
The Killing spinors are now given by
\begin{equation}
\epsilon=[\sqrt{\tilde h^{1/2}+\lambda r^{-1/2}}-
\sqrt{\tilde h^{1/2}-\lambda r^{-1/2}}\,\gamma_1](1+\im \gamma_0)\epsilon_0
\ ,
\end{equation}
where $\tilde h=8g^2Q^2r+\lambda^2r^{-1}$.
Note that these Killing spinors have the same structure as (\ref{eq:7dkil})
corresponding to the domain-wall black hole solution in the $m=0$ limit of
$D=7$, $N=2$ gauged supergravity.

To summarize, we have seen that in addition to the known AdS$_4$ black hole
solutions to (\ref{d4lagxxx}) for non-vanishing $g$ and $m$, there are also
domain-wall black hole solutions in the two distinct limits: $g=0$ with
$k=1$, and $m=0$ with $k=0$.  In both cases, these correspond to new
solutions in the $S^3\times\R^4$ compactification of M-theory.  When $m=0$
the domain-wall black hole is electrically charged in the Cartan subgroup of
the gauge fields arising from $S^3$.  When $g=0$ the charge is instead
carried by a field strength coming from the $\R^4$ reduction.

\subsection{Limit to $S^5\times \R^2$}

    There is a different limit which we may consider, in which three
exponentials, rather than only one, survive in the original
potential in (\ref{d4lagxx}).  To do this, we redefine $\vec\phi$
according to
\be
\vec\phi = \vec\phi' + \vec a_1\, \lambda\ ,
\ee
and then send $\lambda\longrightarrow +\infty$.  The necessary
scalings now imply
\bea
&&g = e^{-\fft12\lambda}\, g'\ ,\qquad
X_1 = e^{-\fft32\lambda}\, X_1'\ ,\qquad
X_m= e^{\fft12\lambda}\, X_m'\ ,\nn\\
&&A_\1^1 = e^{-\fft32\lambda}\, {A_\1^1}'\ ,\qquad
A_\1^m = e^{\fft12\lambda}\, {A_\1^m}'\ ,\nn\\
&&\mu_1 = e^{-\lambda}\, \mu_1'\ ,\qquad \mu_m = \mu_m'\ ,
\eea
where we now split the $i=(1,2,3,4)$ index as $i=1$ and
$i=m=(2,3,4)$.   The scalar potential in (\ref{d4lagxx}) now becomes
\be
V = -4 {g'}^2\,( e^{\phi_1} + e^{\phi_2} + e^{\phi_3})\ .
\ee

     Implementing these scalings in the ans\"atze (\ref{s7metred}) and
(\ref{s7f4red}), we find
\bea
ds_{11}^2 &=& e^{\fft13\lambda}\, \Big\{
\wtd\Delta^{2/3}\, ds_4^2 +g^{-2}\,
\wtd\Delta^{-1/3}\, \Big(X_1^{-1}\, (d\mu_1^2 +\mu_1^2\, d\psi_1^2)\nn\\
&& +
\sum_{m=2}^4 X_i^{-m}\, \Big( d\mu_m^2 + \mu_m^2\,
 (d\psi_m + g\,
A^m_\1)^2 \Big)\Big\}\ ,\label{s7metred3}\\
F_\4 &=& e^{\fft12\lambda} \Big\{ 2g\,\sum_{m=2}^4 \Big(X_m^2\, \mu_m^2 -
\wtd\Delta\, X_m \Big)\,
\ep_\4 +\fft1{2g}\, \sum_{m=2}^4 X_m^{-1}\, {{\bar *}dX_m}\wedge d(\mu_m^2)
\label{s7f4red3}\\
&&- \fft1{2g^2}\, X_1^{-2}\, d(\mu_1^2)\wedge d\psi_1\wedge {\bar *F_\2^1}
-\fft1{2g^2}\, \sum_{m=2}^4 X_m^{-2}\, d(\mu_m^2)\wedge (d\psi_m + g\,
A^m_\1) \wedge
{{\bar *} F_\2^m}\ ,\nn
\eea
where $\sum_{m=2}^4 \mu_m^2=1$.  We have again dropped the primes
after taking the $\lambda\longrightarrow +\infty$ limit.  In this
case, we can recognise the reduction as being on $S^5\times \R^2$.

\subsection{Limits of solutions}

        As we have seen in the previous section, the $S^5\times R^2$
limit is obtained by uniformly shifting the three dilatonic scalars
$\phi_i$.  In fact it is possible to have a consistent truncation of
the Lagrangian (\ref{d4lagxx}) such that all the three scalars are
equal, {\it i.e.}\ $\phi_i = \phi/\sqrt3$, provided that all but one
of the vector potentials are set to zero.  It follows that the
Lagrangian becomes
\be
e^{-1}{\cal L}_4 = R - \ft12 (\del\phi)^2 + 12 m^2\, e^{\phi/\sqrt3}
+ 12 g^2\, e^{-\phi/\sqrt3} - \ft14 e^{\sqrt3\,\phi}\,
F_\2^2\ .
\ee
Note that we have again introduced the additional parameter $m$, by
shifting $\phi$ and rescaling $g$ appropriately.  For non-vanishing
$m$ and $g$, the values of $m$ and $g$ are unimportant, since they can
be changed by a constant shift of the dilaton $\phi$.  The Lagrangian
admits a single charge AdS$_4$ black hole, ({\it i.e.}\ $a=\sqrt3$,)
given by
\bea
ds_4^2 &=& -H^{-1/2}\, f\, dt^2 + H^{1/2}\, f^{-1}
(dr^2 + r^2 (dy_1^2 + dy_2^2))
\ ,\nn\\
f &=& -\fft{\mu}{r} + 4m^2r^2\, H\ ,\qquad e^{\fft2{\sqrt3}\phi}=H\ ,\nn\\
A_\1 &=&\fft{(1+k\sinh^2\alpha)^{1/2}}{\sinh\alpha} H^{-1}\, dt\ ,\qquad
H = \fft{g}{m}\left(1 + \fft{\mu\sinh^2\a}{r}\right)\ .
\eea
As in the previous cases, the solution does not admit an $m=0$ limit,
but it does allow a limit where instead $g\to0$, $\alpha\to\infty$ with
$g\sinh^2\alpha$ held fixed, in which case the solution becomes a
domain-wall black hole.

\section{$S^5$ reduction of type IIB theory, and its limit}

    The Kaluza-Klein reduction of type IIB supergravity on $S^5$ is
expected to admit a consistent truncation to gauged $N=8$ supergravity
in $D=5$, with $SO(6)$ Yang-Mills gauge fields in the supergravity
multiplet.  As usual, the complexity of the reduction procedure has
prevented any complete results from being obtained.  However, in
\cite{ten}, it was shown that a further truncation to $N=2$ supergravity
in $D=5$, with just the $U(1)^3$ Cartan subgroup of the $SO(6)$
Yang-Mills gauge fields surviving, {\it is} explicitly embeddable in
$D=10$ type IIB supergravity.  The ansatz found in \cite{ten}, which
gives a consistent truncation to the $N=2$ supermultiplet comprising
the graviton, three $U(1)$ gauge fields, and two scalar fields is
\bea
ds_{10}^2 &=& \sqrt{\wtd\Delta}\, ds_5^2 + \fft1{g^2\,
\sqrt{\wtd\Delta}}\,  \sum_{i=1}^3 X_i^{-1}\,
\Big(d\mu_i^2 + \mu_i^2\, (d\psi_i +g\, A^i)^2\Big)\ ,\label{2bs5met}\\
G_\5 &=& 2g\, \sum_i\Big(X_i^2\, \mu_i^2 -\wtd\Delta\, X_i\Big)\,
\ep_\5 - \fft1{2g}\, \sum_i X_i^{-1}\, {{\bar *} dX_i}\wedge
d(\mu_i^2) \nn\\
&&+ \fft1{2 g^2} \,
\sum_i X_i^{-2}\, d(\mu_i^2)\wedge (d\psi_i +g\, A_\1^i)
\wedge {{\bar *} F_\2^i}\ .\label{5fs5red}
\eea
where the self-dual 5-form of $D=10$ is given by $F_\5=G_\5+ {*G_\5}$.
The two scalars are parameterised in terms of the three quantities
$X_i$, which are subject to the constraint $X_1\, X_2\, X_3=1$.  They
can be parameterised in terms of two dilatons $\vec\phi=(\phi_1,
\phi_2)$ as
\be
X_i=e^{-\ft12 \vec a_i\cdot \vec \phi}\ ,
\ee
A convenient choice for the dilaton vectors $\vec a_i$ is
\be
\vec a_1 = (\ft{2}{\sqrt6}, \sqrt2)\ , \qquad
\vec a_2 = (\ft{2}{\sqrt6}, -\sqrt2)\ ,\qquad
\vec a_3 = (-\ft{4}{\sqrt6}, 0)\ .
\ee
The three quantities $\mu_i$ are subject to the constraint $\sum_i
\mu_i^2=1$.

    Substituting these ans\"atze into the equations of motion of the
type IIB theory, it was shown in \cite{ten} that one consistently gets
five-dimensional equations of motion that can be derived from the
Lagrangian
\be
e^{-1}\, {\cal L}_5 = R - \ft12(\del\phi_1)^2 -\ft12(\del\phi_2)^2
+ 4g^2\, \sum_i X_i^{-1}- \ft14 \sum_i X_i^{-2}\, (F_\2^i)^2  +\ft14
\ep^{\mu\nu\rho\sigma\lambda}\, F^1_{\mu\nu}\, F^2_{\rho\sigma}\,
A^3_\lambda\ .
\label{d5gauged}
\ee

\subsection{Limit to $S^3\times \R^2$}

    Let us consider a limit under which the dilatonic scalar $\phi_1$
in $\vec\phi=(\phi_1,\phi_2)$ is shifted by a constant $\lambda$,
according to
\be
\phi_1=\phi_1' -\sqrt6\, \lambda\ ,
\ee
with $\phi_2$ left unchanged.  This implies that the quantities $X_i$
will scale as
\be
X_a= e^{\lambda}\, X_a' \ ,\qquad X_3 = e^{-2\lambda}\, X_3' \ ,
\ee
where we have split the $i=(1,2,3)$ index as $i=a=(1,2)$ and $i=3$.
If we also make the scalings
\be
g = e^{-\lambda}\, g'\ ,\qquad A_\1^a = e^{\lambda}\, {A_\1^a}'
\ ,\qquad A_\1^3 = e^{-2\lambda}\, A_\1^3\ ,
\ee
then we can take the limit $\lambda\longrightarrow +\infty$, to obtain
the Lagrangian
\be
e^{-1}\, {\cal L}_5 = R - \ft12(\del\phi_1)^2 -\ft12(\del\phi_2)^2
+ 4g^2\, e^{-\fft2{\sqrt6}\phi_1}
- \ft14 \sum_i X_i^{-2}\, (F_\2^i)^2  +\ft14
\ep^{\mu\nu\rho\sigma\lambda}\, F^1_{\mu\nu}\, F^2_{\rho\sigma}\,
A^3_\lambda\ ,
\label{d5gauged2}
\ee
where we have dropped the primes after taking the limit.

   If we additionally make the rescalings
\be
\mu_a = \mu_a'\ ,\qquad \mu_3 = e^{-\fft32\lambda}\, \mu_3'
\ee
in the reduction ans\"atze (\ref{2bs5met}) and (\ref{5fs5red}), then
we find that they become
\bea
ds_{10}^2 &=& e^{\fft12\lambda}\, \Big\{
\sqrt{\wtd\Delta}\, ds_5^2 + \fft1{g^2\,
\sqrt{\wtd\Delta}}\, \Big( \sum_{a=1}^2 X_a^{-1}\,
(d\mu_a^2 + \mu_a^2\, (d\psi_a +g\, A^a)^2\nn\\
&&\qquad\qquad + X_3^{-1}\, (d\mu_3^2 + \mu_3^2\, d\psi_3^2)\Big) \Big\}
\ ,\label{2bs5met2}\\
G_\5 &=& e^{\lambda}\, \Big\{
2g\, \sum_{a=1}^2\Big(X_a^2\, \mu_a^2 -\wtd\Delta\, X_a\Big)\,
\ep_\5 - \fft1{2g}\, \sum_{a=1}^2 X_a^{-1}\, {{\bar *} dX_a}\wedge
d(\mu_a^2) \label{5fs5red2}\\
&&+ \fft1{2 g^2} \,
\sum_{a=1}^2 X_a^{-2}\, d(\mu_a^2)\wedge (d\psi_a +g\, A_\1^a)
\wedge {{\bar *} F_\2^a}+ \fft1{2g^2}\, X_3^{-2}\, d(\mu_3^2) \wedge
d\psi_3 \wedge {\bar *F_\2^3}\Big\}\ ,\nn
\eea
where, having dropped the primes as usual, we now have $\sum_{a=1}^2
\mu_a^2 =1$, with $\mu_3$ unconstrained, and $\wtd\Delta = \sum_a
X_a\, \mu_a^2$.  We can recognise this as an $S^3\times \R^2$
reduction of the type IIB theory.  The resulting five-dimensional
Lagrangian follows from (\ref{d5gauged}) by applying the limiting
procedure that we have used here, giving
\be
e^{-1}\, {\cal L}_5 = R - \ft12(\del\phi_1)^2 -\ft12(\del\phi_2)^2
+ 4g^2\, e^{-\fft2{\sqrt6}\phi_1}- \ft14 \sum_i X_i^{-2}\, (F_\2^i)^2  +\ft14
\ep^{\mu\nu\rho\sigma\lambda}\, F^1_{\mu\nu}\, F^2_{\rho\sigma}\,
A^3_\lambda\ .
\label{d5gauged3}
\ee

\subsection{New domain-wall black holes in $D=5$}

            In the  $S^5$ limit to $S^3\times R^2$ discussed in the
previous section, we have rescaled the dilaton $\phi_1$, while
$\phi_2$ is left alone.  For simplicity, the dilaton $\phi_2$ can be
consistently truncated out, provided $F^1_{(2)}=F^2_{(2)}=F_{(2)}/\sqrt{2}$.
Thus we shall consider the relevant Lagrangian
\begin{eqnarray}
e^{-1}{\cal L}_5 &=& R - \ft12 (\del \phi)^2 +
8m^2\, e^{\fft1{\sqrt{6}}\phi} + 4g^2 e^{-\fft2{\sqrt{6}}\phi}
-\ft14 e^{\fft2{\sqrt{6}}\phi} (F_\2)^2
-\ft14 e^{-\fft4{\sqrt{6}}\phi} (F^3_\2)^2\nonumber\\
&&+\ft18\epsilon^{\mu\nu\rho\sigma\lambda}F_{\mu\nu}F_{\rho\sigma}
A^3_\lambda\ ,
\end{eqnarray}
which, if we take the limit $m=0$, corresponds to the $S^3\times \R^2$
compactification of the previous subsection.  The $N=2$ fermion supersymmetry
transformations are given by
\begin{eqnarray}
\label{eq:5susy}
\delta\psi_\mu&=&[\nabla_\mu-\ft{\im}{\sqrt{2}} gA_\mu-\ft{\im m^2}{2g}
A_\mu^3 +(\ft{g}3e^{-\fft1{\sqrt{6}}\phi}+\ft{m^2}{6g}
e^{\fft2{\sqrt{6}}\phi}) \gamma_\mu\nonumber\\
&&\qquad\qquad\qquad
+\ft{\im}{24}(\gamma_\mu{}^{\nu\lambda}
-4\delta_\mu^\nu\gamma^\lambda)
(e^{\fft1{\sqrt{6}}\phi}F_{\nu\lambda}+
e^{-\fft2{\sqrt{6}}\phi}F^3_{\nu\lambda})]\epsilon\ ,\\
\delta\lambda&=&[-\ft{\im}{4}\gamma^\mu\partial_\mu\phi
-\ft{\im}{\sqrt{6}}(ge^{-\fft1{\sqrt{6}}\phi}
-\ft{m^2}{g} e^{\fft2{\sqrt{6}}\phi})+\ft1{8\sqrt{6}}
(e^{\fft1{\sqrt{6}}\phi}F_{\mu\nu}
-2e^{-\fft2{\sqrt{6}}\phi}F^3_{\mu\nu})\gamma^{\mu\nu}]\epsilon\ .\nonumber
\end{eqnarray}
This Lagrangian admits an AdS$_5$ black hole solution, given by
\bea
ds_5^2 &=& -H^{4/3}\, f\, dt^2 + H^{2/3}(f^{-1}\, dr^2 +
r^2\,  (dy^i dy_i))\ ,\nn\\
f&=& -\fft{\mu}{r^2} + m^2 r^2 H^2\ ,\qquad e^{\sqrt6 \phi} = H\ ,
\nn\\
A_\1 &=& \fft{\sqrt{2}(1+k\sinh^2\alpha)^{1/2}}{\sinh\a}\, H^{-1} dt\ ,\qquad
A^3_{(1)}=0\ ,\qquad
H = \fft{g}{m}\left(1 + \fft{\mu\sinh^2\a}{r^2}\right)\ .\label{someeq}
\eea
Note that this solution was obtained in \cite{Behrndt2} for $m=g$.
When $k=0$, it can also be obtained \cite{Cveticgubser1,ten} from the
$S^5$ reduction of the rotating D3-brane \cite{klt,rosf,ten}.  It follows
from (\ref{someeq}) that the solution does not have an $m=0$ limit, but
it does have a $g\to0$ limit with $g\sinh^2\alpha$ held fixed, which gives
rise to a domain-wall black hole solution with $k\ne 0$.

          When $m=0$, the Lagrangian fits the general pattern of the
Lagrangian (\ref{gensinglelagsp}) in the appendix.  Thus in this limit,
we can find a new domain-wall black hole solution, given by
\bea
ds_5^2 &=& -f\, dt^2 + f^{-1} dr^2 + r\, (dy_1^2 + dy_2^2 + dy_3^2)
\ ,\qquad e^{\fft{2}{\sqrt6}\phi} = r\ ,\nn\\
f &=& 2r\, (\ft89 g^2 + \fft{\lambda^2}{4r^3} + \fft{\mu}{r^{3/2}})
\ ,\qquad A_\1 = \lambda\, r^{-3/2}\, dt\, \qquad A_{(1)}^3=0\ .
\eea
This solution preserves half of the $N=2$ supersymmetry provided $\mu=0$.
As in the $D=7$ and the $D=4$ cases, we find that
the Killing spinors are given by
\begin{equation}
\epsilon=[\sqrt{h^{1/2}-\ft1{\sqrt{2}}\lambda r^{-1}}
-\sqrt{h^{1/2}+\ft1{\sqrt{2}}\lambda r^{-1}}\,\gamma_1](1+\im \gamma_0)
\epsilon_0\ ,
\end{equation}
where $h=\fft{16}{9}g^2r+\fft12\lambda^2r^{-2}$.  This $m=0$ solution,
with $k=0$, corresponds to a domain-wall black hole of the $S^3\times
\R^2$ compactification obtained in the previous subsection.

\section{Conclusions}

We studied in detail a class of consistent Kaluza-Klein sphere
compactifications of string/M-theory whose effective theories reduce to
gauged supergravity theories that do not admit anti-de Sitter (AdS)
space-time as a vacuum solution.  We refer to them as domain-wall
supergravities. This class of supergravity solutions can be viewed as
particular limits of the (standard) AdS gauged supergravity theories in
$D=7,4,5$, which are obtained as Kaluza-Klein compactifications
on $S^4$ and $S^7$ of eleven-dimensional supergravity and on $S^5$ of
ten-dimensional Type IIB supergravity.

We obtained specific limits of AdS gauged supergravity theories by taking
certain moduli, associated with the non-homogeneous deformations of
$S^n$, to their boundary values, which result in compactifications of
string/M-theory on $S^p\times R^q$ ($p+q=n$) spaces and correspond to
domain-wall supergravities. We classified such possible limits for the
abelian truncations of the gauged AdS supergravities in $D=7,4,5$.
A particular instructive example is a limit of M-theory compactifications
on $S^{4}$ which reduces to a compactification on $S^{3}\times R$,
and a domain-wall supergravity in $D=7$. This compactification can
be reinterpreted as a consistent compactification of Type IIA theory
on $S^3$.  The limiting procedure we employed highlights a geometrical
interpretation of the massive, but ungauged supergravity in $D=7$ and
the massless $D=7$ gauged supergravity.

Since the sphere compactifications of the AdS supergravities are believed
to be consistent, it follows that the resulting domain-wall supergravities
are consistent sphere compactifications of higher dimensional supergravities
as well.  While these domain-wall supergravities were obtained as
certain limits of AdS supergravities, they now stand as valid sphere
compactifications in their own right.

Typical solutions of domain-wall supergravities are of the
(black-hole) domain-wall-type, \ie asymptotically the
configurations have non-constant scalar fields. We have shown how some
of these solutions can be obtained as limits of the existing (black-hole)
solutions of the corresponding AdS gauged supergravities.  However, we
have also found new classes of domain-wall black holes that cannot be
obtained as special limits of the AdS supergravity solutions and have
shown that they are supersymmetric.  These sets of solutions may play an
important role in the study of the dual quantum field theories (QFT),
according to the proposed domain-wall/QFT correspondence~\cite{BST}.

\bigskip\bigskip

\noindent{\bf Note added}:

  After this work was completed, a paper appeared in which the complete
non-linear ansatz for the $S^4$ reduction of $D=11$ supergravity was
presented \cite{nvvn}.

\section*{Appendix}

\appendix
\section{A class of domain-wall black holes}

      Let us consider a bosonic system with the metric, a vector
potential and a dilaton, with the Lagrangian given by
\be
e^{-1} {\cal L} = R- \ft12 (\del\phi)^2 -\ft14 e^{a\phi}\, F_\2^2 -
V(\phi)\ ,\label{gensinglelag}
\ee
where $V(\phi)$ is the scalar potential.  We consider the following
ans\"atze
\bea
ds^2 &=& -e^{2u}\, dt^2 + e^{2A}\, dy^i dy^i + e^{2v}\, dr^2\ ,\nn\\
A_\1 &=& w\, dt\ .
\eea
where $u$, $v$, $w$, $C$ and the dilaton $\phi$ are
functions of $r$ only.   The dimension of the space $dy^idy^i$ is
$p=D-2$.  The equations of motion is then given by
\bea
u'' +u'(u'-v' + pA') &=& \fft{p-1}{2p} w'^2\, e^{a\phi - 2u} -
\fft{1}{p}\, V\, e^{2v}\ ,\nn\\
A'' + A'(u'-v'+pA') &=& - \fft{1}{2p} w'^2\, e^{a\phi-2u}
-\fft{1}{p}\, V\, e^{2v}\ ,\nn\\
A'' -A'(u' + v' -A') &=& -\fft{1}{2p} \phi'^2\ ,\label{eom}\\
\phi'' + \phi'(u'-v' + pA')&=& -\ft12a\, w'^2\, e^{a\phi -2u} +
V_{,\phi}\, e^{2v}\ ,\nn\\
(w'\, e^{a\phi -u - v + pA})' &=& 0 \ ,\nn
\eea
where a prime denotes a derivative with respect to $r$.  This set of
equations are self-consistent, implying the existence of solutions.

        In this paper, we showed that there exists a modulus limit for
generic gauged supergravities, where the relevant part of the Lagrangian
has the form (\ref{gensinglelag}) with
\be
V(\phi) = -\ft12 g^2 e^{-a\phi} \qquad
{\rm and} \qquad a^2 =\fft{2}{p}\ .
\ee
In other words, the Lagrangian has the form
\be
e^{-1} {\cal L} = R- \ft12 (\del\phi)^2 -\ft14 e^{\sqrt{2/p}\,\phi}\,
F_\2^2 + \ft12 g^2\, e^{-\sqrt{2/p}\,\phi}\ .\label{gensinglelagsp}
\ee
In this case, the equations of motion (\ref{eom}) admit a closed-form
solution, given by
\bea
ds^2 &=& -f\, dt^2 + f^{-1}\,  dr^2 + r\,  dy^i dy^i\ ,\nn\\
f&=& 2r\, \Big(\fft{g^2}{p^2} + \fft{\lambda^2}{4r^p} +
\fft{\mu}{r^{p/2}} \Big)
\ ,\qquad A_\1 = \lambda \, r^{-p/2}\, dt\ ,\qquad
e^{\sqrt{2/p}\,\phi} = r \ .\label{gensolsp}
\eea
The solution describes a domain-wall black hole, whose geometry  approaches
a pure domain-wall spacetime as $r\rightarrow \infty$.

\end{document}